\documentclass{article}

\usepackage{microtype}
\usepackage{graphicx}
\usepackage{subfigure}
\usepackage{booktabs}

\usepackage{hyperref}

\usepackage[accepted]{icml2019}

\usepackage{amsmath}
\usepackage{mathrsfs}
\usepackage{amsfonts}
\usepackage{amsthm}
\usepackage{dialogue}
\newcommand{\cX}{\mathcal{X}}

\newcommand{\cU}{\mathcal{U}}

\newcommand{\bR}{\mathbb{R}}

\newtheorem{prop}{Proposition}

\newcommand{\xt}{x_{1:t}}

\newcommand{\cbar}{\, | \,}

\newcommand{\csemi}{\, ; \,}

\newcommand{\eqnref}[1]{(\ref{eqn:#1})}

\def \tq {\tilde q}

\usepackage[title]{appendix}

\icmltitlerunning{Shaping the Narrative Arc}

\begin{document}

\twocolumn[
	\icmltitle{Shaping the Narrative Arc: An Information-Theoretic Approach to Collaborative Dialogue}
	\icmlsetsymbol{equal}{*}

	\begin{icmlauthorlist}
	\icmlauthor{Kory W.~Mathewson*}{uofa}
	\icmlauthor{Pablo Samuel Castro}{goob}
	\icmlauthor{Colin Cherry}{goot}
	\icmlauthor{George Foster}{goot}
	\icmlauthor{Marc G.~Bellemare}{goob}
	\end{icmlauthorlist}

	\icmlaffiliation{uofa}{Department of Computing Science, University of Alberta, Edmonton, Canada, *during internship with Google Brain.}
	\icmlaffiliation{goob}{Google Brain, Montreal, Canada}
	\icmlaffiliation{goot}{Google Translate, Montreal, Canada}

	\icmlcorrespondingauthor{Kory Mathewson}{korymath@gmail.com}
	\icmlkeywords{Machine Learning, ICML}

	\vskip 0.3in
]

\printAffiliationsAndNotice{}  %

\begin{abstract}
We consider the problem of designing an artificial agent capable of interacting with humans in collaborative dialogue to produce creative, engaging narratives. In this task, the goal is to establish universe details, and to collaborate on an interesting story in that universe, through a series of natural dialogue exchanges. Our model can augment any probabilistic conversational agent by allowing it to reason about universe information established and what potential next utterances might reveal. Ideally, with each utterance, agents would reveal  just enough information to add specificity and reduce ambiguity without limiting the conversation. We empirically show that our model allows control over the rate at which the agent reveals information and that doing so significantly improves accuracy in predicting the next line of dialogues from movies. We close with a case-study with four professional theatre performers, who preferred interactions with our model-augmented agent over an unaugmented agent.
\end{abstract}

\section{Introduction} \label{introduction}

Designing and building computational models that generate meaningful dialogue for human-interaction, in an interesting and engaging manner, is a challenging open problem. As personal digital assistants increase in popularity, proper conversational capabilities may allow them to provide creative, playful, and helpful interactions. Conversational agents can be effective for health-care \cite{BICKMORE2006556}, by supporting cognitive-behavioral therapy for treating depression \cite{fitzpatrick17delivering}, helping patients with chronic pain \cite{doi:10.1001/jama.2017.14151}, and supporting reminiscence \cite{nikitina2018smart}. These applications require systems capable of understanding and collaboration.

What makes some dialogues more interesting than others? Interesting collaborative dialogue constructs knowledge iteratively \cite{swain2000output} and depends on each speaker bringing information to the conversation \cite{sawyer2003improvised}.
Interestingness is also subjective and difficult to directly optimize via numerical methods \cite{2016arXiv160601541L,venkatesh2018evaluating}.

Rule-based conversational models have existed for over 50 years \cite{weizenbaum1966eliza}. These methods are limited by hand-tuning and engineering to predict and handle possible inputs. Generative language models maximize the likelihood of an utterance (e.g. a sentence or sequence of words) \cite{graves2013generating}. These models can predict the likelihood of an utterance by considering the sentences as a sequences of words, sub-word units, characters, and/or tokens \cite{sennrich2015neural}. This objective can result in generated sentences which are grammatically correct, and bear a semantic relationship to the context surrounding it, but lack global consistency \cite{liu2018generating}.

Our work generates interesting dialogue by using a narrative arc to incrementally construct shared knowledge. A narrative arc defines evolving qualities of emotion, tension, or topic over a story \cite{bizzocchi2007games}. We draw inspiration from improvised theater, where actors collaborate in real time to develop narrative based on thematic constraints \cite{johnstone1979impro}. Improvised theater is a unique storytelling medium which relies on collaborative dialogue in which each utterance must carry significant information \cite{swain2000output}. We appeal to the two \textit{golden rules} of improvised dialogue: 1) \textbf{accept} (i.e. be consistent with the dialogue thus far and 2) \textbf{reveal} (i.e. progress the dialogue with new information) \cite{sawyer2003improvised,johnstone1979impro}.

In this work, we propose a new method to modulate a conversation model, which \textbf{accepts} input utterances by generating consistent and \textbf{revealing} responses. Our approach combines a conversational model with a topic classifier, or \emph{universe model}. We borrow the term \emph{universe} from improvised theater where it is used to describe the \textit{world-as-we-know-it} \cite{johnstone1979impro,mcleod2000use,raby2010improvisation}. The universe encompasses associations surrounding the dramatic world, and is motivated by the possible world semantics theory \cite{kripke1963semantical}.

We identify two modes of operation for our shaping method: \emph{revealing} and \emph{concealing}. Revealing dialogue adds additional information about the current universe. Generating utterances which progress a scene with new information is the primary goal of our approach. Concealing dialogue avoids exposing new information about the universe. 

The universe model characterizes the information revealed by each utterance in a sequence. We refer to this information profile across utterances as the narrative arc. By tuning the revealingness we can selectively choose utterances to shape the narrative arc to produce more interesting and engaging dialogue. We argue that a balance between revealing and concealing is required for interesting and engaging collaborative dialogue \cite{swain2000output}. Both over-specification and ambiguity are undesirable \cite{sawyer2003improvised,johnstone1979impro}. We hypothesize that there is an ideal region of information revelation which our method can expose in existing text-based narratives such as movie scripts.

\section{Shaping the Narrative Arc} \label{sop}
In this section, we present a mechanism for shaping the narrative arc inspired by combining methods exploring entropy in textual documents \cite{shannon1951prediction} with the \textit{Simple Shapes of Stories} described by Vonnegut.\footnote{From K. Vonnegut lecture: goo.gl/JuEDVR} We describe concepts of conversation and universe models. Then, we show how these combine to describe a narrative arc. Finally, we show how the narrative arc can be used to generate interesting dialogue.

\subsection{The Conversation Model}
\label{sec:conversation_model}

A conversation model accepts an input utterance and generates one, or several, output utterance(s). The conversation model maintains local coherence by conditioning output generation on the input. We write $\cX$ to denote the set of possible utterances (i.e. sequences of words); in this work, $\cX$ is a collection of English sentences. A sequence of $t$ successive utterances is a \emph{dialogue}, denoted $\xt$. A conversation model yields a probability $q$ of an utterance $x_t$ given a dialogue $x_{1:t-1}$.

We focus on dialogue generation using three retrieval-based conversation models. The first two models are based on the OpenSubtitles dataset \cite{lison2018opensubtitles}. Pre-processing details are included in the supplementary material. When queried with an input line $x_{t-1}$, a model returns $K$ candidate responses:

\begin{itemize}
  \setlength\itemsep{-0.2em}
  \item {\bf Baseline Random model}: sample $K$ lines from $\cX$.
  \item {\bf Deep neural network model (DNN)}: we embed all the lines in $\cX$ into a latent semantic space $S$ using the Universal Sentence Encoder \cite{cer2018universal}. We encode the input line into $S$, and return the $K$ approximate nearest neighbours in $S$ using the $L^2$ norm as the distance metric.
  \item \textbf{Books}: Similar to the DNN model, responds with semantically-related nearest neighbor lines from literature, filtered for offensive content.\footnote{books.google.com/talktobooks}
\end{itemize}

\subsection{The Universe Model} \label{sec:universe_model}

The universe model measures how each successive utterance of a dialogue influences the probability distribution over universes. For a given utterance, the universe model calculates a probability distribution over universes. For a sequence of utterances, we use recursive universe belief propagation (Sec. \ref{sec:rubp}) to update the posterior over the course of a dialogue. Revealing dialogue would concentrate probability mass on a single universe, and concealing dialogue would distribute probability mass uniformly across a set of universes. The shape of this sequence of posteriors is the narrative arc (Sec. \ref{sec:arc}).
We investigated reveal/conceal dynamics using three different universe models based on probabilistic topic classifiers.
\begin{itemize}
  \setlength\itemsep{-0.2em}
  \item {\bf Newsgroups}: Using the newsgroup classification dataset, we filter out stop-words, created frequency vectors, and use the TF-IDF (term frequency / inverse document frequency) \cite{salton1988term} word weighting scheme to account for word importance in the corpus. We train a na\"{i}ve Bayes classifier on 5 aggregate topic universes (\textsc{Computers}, \textsc{Recreation}, \textsc{Religion}, \textsc{Science}, and \textsc{Talk}) \cite{joachims1996probabilistic}.
  \item {\bf Movies}: Na\"{i}ve Bayes classifier, trained similar to Newsgroups, using a collected dataset of film synopses and one of 10 corresponding genres (\textsc{Drama}, \textsc{Comedy}, \textsc{Horror}, \textsc{Action}, \textsc{Crime}, \textsc{Romantic Comedy}, \textsc{Romance}, \textsc{Thriller}, \textsc{Film Adaptation} and \textsc{Silent Film}) from Wikipedia data \cite{hoang2018predicting}.
  \item {\bf DeepMoji}: Deep neural network that takes input text and outputs a distribution over a set of 8 aggregated emoji universes: (\textsc{Sad}, \textsc{Mad}, \textsc{Meh}, \textsc{Nervous}, \textsc{Glad}, \textsc{Music}, \textsc{Love}, and \textsc{Miscellaneous}) \cite{2017arXiv170800524F}. The authors' pretrained model was used.\footnote{github.com/bfelbo/DeepMoji}
\end{itemize}

\begin{figure*}[ht]
  \centering
    \includegraphics[width=0.8\textwidth]{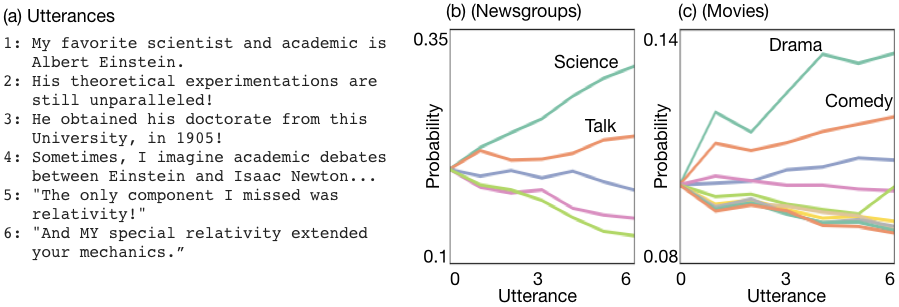}
  \caption[Example of a Narrative Arc]{The narrative arcs of a synthetic dialogue (a), using the Newsgroups universe model (b) and Movies universe model (c). This dialogue is likely \textsc{SCIENCE} or \textsc{TALK} under the Newsgroups model, and \textsc{DRAMA} or \textsc{COMEDY} under the Movie genres model.}
  \label{fig:example_arc}
\end{figure*}

\subsection{Recursive Universe Belief Propagation}
\label{sec:rubp}

We desire a means by which we can update the universe belief incrementally as evidence is accumulated with each successive utterance in a dialogue. We begin by defining the notion of a \emph{universe model} as a means of modelling the dynamics of information revelation. Consider a finite set of \emph{universes}, $\cU$. The role of a universe model is to assess the compatibility of an utterance with a given discrete universe, $u\in\mathcal{U}$. Given such a model, we develop a method to update the agent's posterior universe distribution over a sequence of utterances. For each universe $u$, the universe model assigns a likelihood $p(x_t ~|~ x_{1:t-1}, u)$ to an utterance $x_t$, conditioned on a dialogue $x_{1:t-1}$.

The universe model iteratively updates a posterior belief over universes, in a similar spirit to prediction with expert forecasters \cite{cesabianchi06prediction}. The probability of a given universe depends on iteratively combining evidence in support of that universe. We define the the posterior probability over universes $\cU$ given a sequence of $t$ utterances $x_{1:t}$ as:
\begin{align*}
p_{t}(u~|~x_{1:t}) &= p_{t-1}(u~|~x_{1:t-1}) \times \frac{p(x_t~|~x_{1:t-1}, u)}{p(x_t~|~x_{1:t-1})}
\end{align*}
Where $p_{t-1}(u|x_{1:t-1})$ is the prior probability, $p(x_{t}|x_{1:t-1}, u)$ is the likelihood of utterance conditioned on the past dialogue and universe, and $p(x_{t}|x_{1:t-1})$ is the likelihood of utterance under the conversation model.

Let $p_0(u|\cdot) = 1 / |\cU|, u \in \cU$ be an initially uniform distribution over universes, or the universe model's prior. We can marginalize out the universe if the evidence is consistent over all hypotheses. To illustrate the relationship between utterance likelihood and universe, we can explicitly write the marginal likelihood as:
\begin{align*}
p(x_t ~|~ x_{1:t-1}) = \sum_{u'} p_{t-1}(u'~|~x_{1:t-1}) p(x_{t}~|~x_{1:t-1}, u')
\end{align*}
Thus, the posterior is updated recursively as:
\begin{align}
\label{eqn:posterior_fn}
p_t(u~|~x_{1:t}) &= p_{t-1}(u~|~x_{1:t-1}) \nonumber \\
&\times \frac{p(x_t ~|~ x_{1:t-1}, u)}{\sum_{u'} p_{t-1}(u'~|~x_{1:t-1}) p(x_t ~|~ x_{1:t-1}, u')}
\end{align}
In practice, it may be convenient to use the output $z(u|x_t)$ of a probabilistic classifier \emph{in lieu} of a likelihood function conditioned on past utterances $x_{1:t}$ and universe $u$. Universe classifiers can be trained separately from language models, and provide a complementary signal if model input distributions overlap. This assumption is justified when both models work with similar training corpus vocabularies. We view the probability distribution over universes output by the universe model as derived from a joint distribution $z(u, x_t)$, of the universe $u$, and utterance $x_t$. With $z(u)$ as the prior distribution over universes, the conditional probability is:
\begin{equation*}
z(u ~|~ x_t) = \frac{z(u, x_t)}{z(x_t)} =  z(u) \times \frac{z(x_t ~|~ u)}{z(x_t)}
\end{equation*}
We can substitute $z(\cdot|x_t)$ for $p(x_t|x_{1:t-1}, \cdot)$ in Eq. \ref{eqn:posterior_fn} by assuming conditional independence (i.e., ${p(x_t~|~x_{1:t-1},u) = p(x_t~|~u)}$), uniform prior distribution (i.e., ${z(u) = 1 / |\cU|, u \in \cU}$) and constant marginal probability (i.e., ${z(x_t) = \sum_{u'} p_{t}(u') p(x_t ~|~ u')}$). These assumptions are justified when the probabilistic topic classifier is a na\"{i}ve Bayes classifier \cite{bishop06pattern} with uniform prior. Thus, the substitution follows the following steps:
\begin{align*}
p(x_t|x_{1:t-1}, u) &\approx z(x_t|u) && \text{[cond. independence]}\\
&= \frac{z(u|x_t)z(x_t)}{z(u)} && \text{[Bayes' theorem]}\\
&\approx z(u|x_t)z(x_t) && \text{[$z(u)$ uniform prior]}\\
&\approx z(u|x_t) && \text{[$z(x_t)$ const. marginal]}
\end{align*}
Eq. \ref{eqn:posterior_fn} thus becomes:
\begin{align}
\label{eq:weighting}
p_t(u~|~x_{1:t}) =&~ p_{t-1}(u~|~x_{1:t-1}) \nonumber \\
&\times \frac{z(u~|~x_t)}{\sum_{u'} p_{t-1}(u'~|~x_{1:t-1}) z(u' ~|~ x_t)}
\end{align}

\subsection{The Narrative Arc} \label{sec:arc}

As defined in Eq. \ref{eq:weighting}, the posterior $p_t(\cdot)$ is a function of the dialogue $\xt$. We define the \emph{narrative arc} as the sequence of universe distributions $p_0(\cdot), p_1(\cdot), \dots$ iteratively calculated for the dialogue. The arc depicts the evolution of a belief over a set of universes. The \emph{narrative arc function} maps $\cX^t \to \mathcal{S}(\cU)^t$, where $\mathcal{S}(\cU)$ is a probability simplex over $\cU$. We discuss three properties of the narrative arc of the synthetic dialogue shown in Fig. \ref{fig:example_arc}:

\textbf{1. Utterances affect the arc in varying degrees.} ``My favorite scientist and academic is Albert Einstein'' is similarly likely under \textsc{Science} and \textsc{Talk}, and less likely under the \textsc{Recreation} universe (bottom green line). Different utterances should have different effects on $p_t(\cdot)$.

\textbf{2. A concentrating posterior signals a revealing dialogue.} A dialogue which emphasizes scientific content, for example, should see $p_t(\textsc{Science}|\cdot) \to 1$. Conversely, we would expect a concealing dialogue to spread the posterior across multiple universes.

\textbf{3. A universe model is a perspective on dialogue.} Different universe models can expose different aspects of the same dialogue. Replacing the Newsgroups universe model by a Movies universe model suggests the dialogue is from a \textsc{Drama} and/or \textsc{Comedy} universe. This dialogue would be considered revealing under both universe models.

\begin{figure*}[ht!]
  \centering
  \includegraphics[width=.9\textwidth]{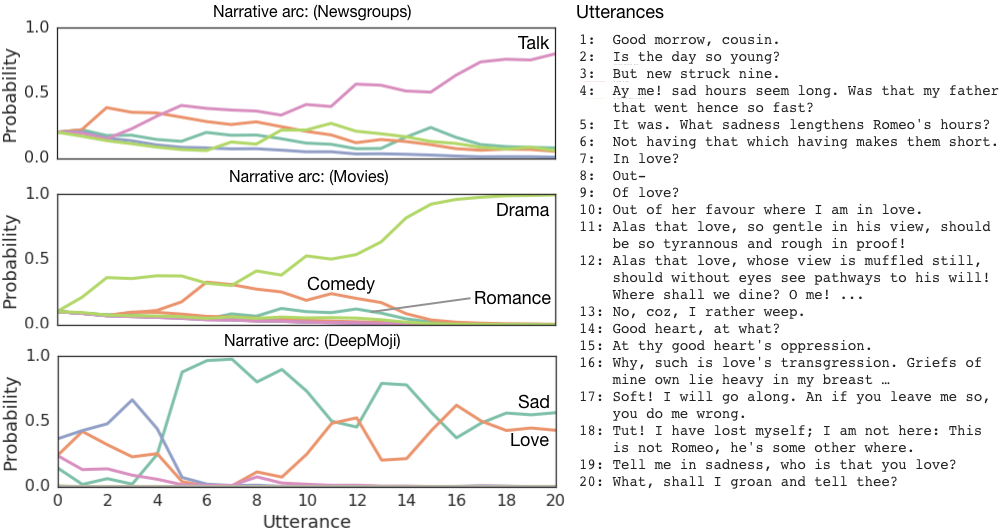}
  \caption{First 20 lines of Romeo and Juliet modeled with Newsgroups (top), Movies (middle), and DeepMoji (bottom) universe models.}
  \label{fig:randj-posterior}
\end{figure*}

In this section, the universe model was applied to a fixed dialogue $\xt$, but the model also provides a criterion for favoring utterances when \emph{generating} dialogue.

\subsection{Generating Dialogue with the Narrative Arc} \label{sec:dialogue}
The entropy of the posterior $p_t(\cdot)$ is given by:
\begin{equation*}
  H(p_t(\cdot)) := -\sum_{u \in \cU} p_t(\cdot) \log p_t(\cdot)
\end{equation*}
Then, the entropy change $\Delta(\cdot)$ due to a new utterance, $x_t$, given the past dialogue, $x_{1:t-1}$, is defined as:
\begin{equation*}
  \Delta(x_t \csemi x_{1:t-1}) := H(p_{t-1}(\cdot)) - H(p_t(\cdot))
\end{equation*}
The term $\Delta(x_t \csemi x_{1:t-1})$ measures how much a given utterance $x_t$ changes the entropy of the posterior, given the previous utterances $x_{1:t-1}$. A positive value of $\Delta(\cdot)$ is a reduction in entropy (i.e. revealing). Conversely, a negative value of $\Delta(\cdot)$ is an increase in entropy (i.e. concealing). We define the \emph{score} of an utterance $x_t$, with respect to a dialogue, $x_{1:t-1}$, as:
\begin{equation*}
  \sigma(x_t \csemi x_{1:t-1}) := \exp \{ \alpha \Delta(x_t \csemi x_{1:t-1}) \}, \quad \alpha \in \bR
\end{equation*}
The exponential function is a convenient way to ensure strict positivity and preserve the ordering of scored candidates.
We use our entropy-based score function $\sigma$ to modulate the sampling of a base conversation model, $q$, toward $\tq$, which depends on the change in entropy due to the new utterance.
\begin{equation}
  \label{eqn:tq}
	\tq(x_t \cbar x_{1:t-1}) \propto q(x_t \cbar x_{1:t-1}) \times \sigma(x_t \csemi x_{1:t-1})
\end{equation}
If $\alpha = 0$, $\sigma(\cdot) = 1$ and candidates are sampled according to $\tq = q$. If $\alpha \neq 0$, $q$ is modulated by the score $\sigma(\cdot)$. 
Modulation \emph{mode} depends on the value of $\alpha$:
\begin{itemize}
	\setlength\itemsep{-0.2em}
	\item $\alpha > 0$ (\textbf{reveal}): modulate $q$ towards revealing the universe. The probability of utterances likely under the universe with highest probability are increased.
	\item $\alpha < 0$ (\textbf{conceal}): modulate $q$ towards concealing the universe. The probability of utterances likely under multiple unlikely universes is increased. Utterances not supporting the likely universe are made more likely.
\end{itemize}

We use these two modulations for filtering samples from our base conversation model. We filter via one of two methods for sampling from an unnormalized distribution: {\bf greedy sampling} and {\bf rejection sampling}. Greedy sampling scores a set of samples from the conversation model and selects the candidate with the maximum score. Scoring a large set of candidates can be time intensive. Rejection sampling (Alg. \ref{alg:rejection}) can sample from the desired unknown modulated distribution online \cite{murphy2012prob}. Additional details on rejection sampling are included in the supplementary material. As the entropy function is bounded, the utterance score $\sigma$ is bounded. In practice, we set a max score and weigh all utterance scores $\sigma$ above the threshold equally. Both filtering methods have benefits. Rejection sampling provides a smoother distribution and does not require scoring a large set of candidates. Greedy sampling is less sensitive to the range of $\Delta$ from different utterances.

\begin{algorithm}[htb]
  \begin{algorithmic}
    \STATE \textbf{Given:} conversation model $q$, scoring function $\sigma$, first line $x_1$, length $N$, max score $M$, max samples $S$
    \STATE \textbf{Return:} dialogue $x_{1:N}$
    \FOR{$t$ in $2...N$}
      \WHILE{step $\le$ $S$}\STATE{
        sample $x_t \sim q(x_t|x_{1:t-1})$ \\[1pt]
        sample $r \sim \text{Uniform}(0, 1)$\;
        }
          \IF{$r \le \sigma(x_t \csemi x_{1:t-1}) / M$}
          \STATE{
            append $x_t$ to $x_{1:t-1}$ \\[1pt]
            break\;
          }
          \ENDIF
        \ENDWHILE
    \ENDFOR
  \end{algorithmic}
 \caption{Generating dialogue with rejection sampling.
 }
 \label{alg:rejection}
\end{algorithm}

\section{Evaluation} \label{sec:evaluation}

\subsection{Narrative Arc of Existing Dialogues} \label{sec:existing}
In Fig. \ref{fig:randj-posterior}, we visualize the narrative arc underlying the first 20 lines of Shakespeare's Romeo and Juliet using three universe models: 1) Newsgroups, 2) Movies, and 3) DeepMoji.

Fig. \ref{fig:randj-posterior} illustrates the entropy-reducing nature of good dialogue by showing us the underlying, evolving, narrative arc. Under the \textit{Newsgroups} universe model, the dialogue evolves toward a \textsc{talk}-centric universe. Under the \textit{Movies} model, the same dialogue balances between comedy and drama before shifting towards drama. Finally, using the \textit{DeepMoji} universe model, a developing ambiguity between DeepMoji universes \textsc{Sadness} and \textsc{Love} is uncovered. This supports the hypothesis that existing dialogues exhibit underlying narrative arcs conditioned on universe models. Additional samples exposing narrative arc dynamics are presented in the supplementary material.

\begin{figure*}[ht]
  \centering
  \includegraphics[width=.9\textwidth]{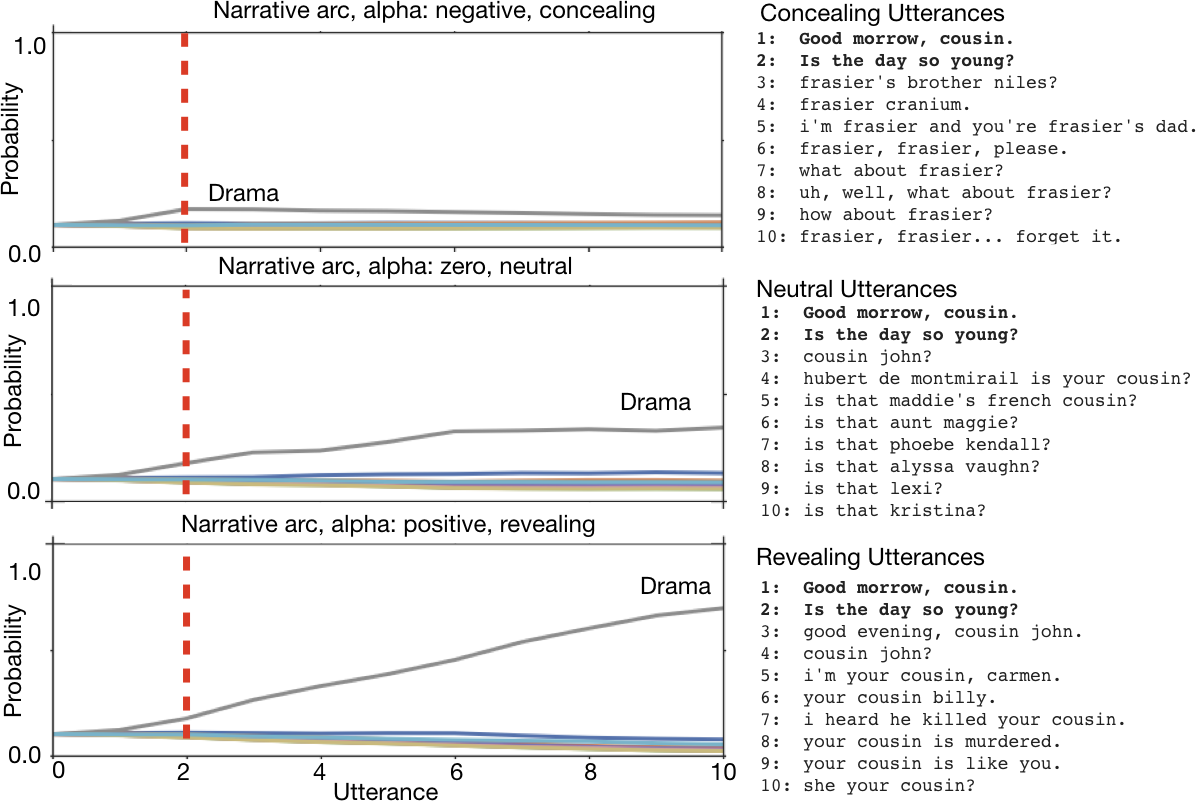}
  \caption{Narrative arcs over 10 utterances at increasing $\alpha$ values:  concealing (top), neutral (mid), revealing (bottom). On the right are utterances generated by each model after priming (bold). Dotted red line indicates the start of narrative arc shaping.
  }
  \label{fig:alpha-posterior}
\end{figure*}

\subsection{Shaping the Narrative Arc} \label{sec:fitting}
In this section, we demonstrate that our method is able to modulate conversation models toward generation of revealing or concealing dialogues. Linguistic quality and semantic consistency of utterances are determined by the language underlying the conversation model. Here, we emphasize evaluation of narrative arc shaping.

We use the DNN conversation model
to test how preferential selection, induced by our score function, can modulate information introduced into the conversation. In Fig. \ref{fig:alpha-posterior} we present characteristic narrative arcs and dialogues using concealing (top), neutral (middle), and revealing (bottom) modes. Each generation was primed with the first two lines from Romeo and Juliet (shown in bold in Fig. \ref{fig:alpha-posterior}).

A significant difference is exposed between concealing (top) which tends toward a high entropy, uniform universe distribution, and revealing (bottom) where drama tends toward $1.0$. \textsc{Drama} remains the most likely universe (and visible on all plots) as it was supported by the first two lines and subsequent utterances did not significantly shift the distribution. Fig. \ref{fig:alpha-posterior} also shows the dialogue generated by the model. Concealing utterances do not add information to the dialogue, revealing utterances incorporate new information over the course of the dialogue.

We next evaluate our method's ability to generate concealing/revealing dialogue by measuring the entropy under both an objective universe (i.e. the universe model used for scoring in generation) and a test universe not used for scoring. We use the Newsgroups universe model for objective scoring and the Movies model for testing. A random conversation model is used to generate response candidates.

We generate 20 conversations following a process similar to Algorithm \ref{alg:rejection} but using greedy sampling.
Each conversation starts with a random dialogue starter line to encourage diversity and then 19 lines are sampled from the conversation model using the narrative arc function. This approximates the length of a medium-duration improvised conversation \cite{sawyer2003improvised}.

\begin{figure}[ht]
  \centering
  \includegraphics[width=0.48\textwidth]{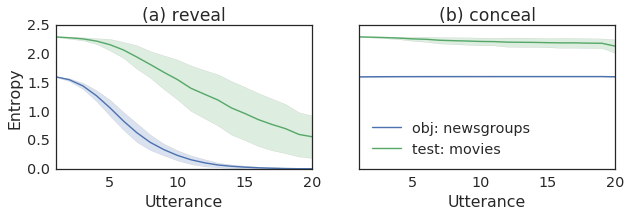}
  \caption{Revealing and Concealing across Universe Models. Dialogue generated to be (a) revealing ($\alpha = 20$) under the objective model Newsgroups is revealing under the testing Movies universe. The same is true for (b) concealing ($\alpha = -25$) dialogue. Data shown are means and standard deviation (shaded) over 20 runs of random conversation model.}
  \label{fig:multiverse}
\end{figure}

Results are presented in Fig. \ref{fig:multiverse}. There is a significant difference between the entropy under the objective and testing universes, but each model exhibits similar dynamics over the dialogues. We conclude that concealing dialogue can conceal under multiple universes, and revealing dialogue can reveal information under multiple universe models.

The revealing/concealing dynamics of each utterance may be related to measurable lexicographical qualities such as words per sentence (WPS). We analyzed the language used in $190$ lines from each model and found a significant difference ($p < 0.001$) between utterances selected by the revealing model ($9.26 \pm 5.7$ WPS) and utterances selected by the concealing model ($5.05 \pm 2.79$ WPS).
\subsection{Predicting the Next Best Line}
\label{arc:sec:interactive}
We next test the system's ability to add information to improve performance on a prediction task. Given a sequence of $5$ gold-standard conversational utterances and a list of $10$ next utterance candidates (i.e. the ground truth and $9$ distractors), can the universe model be used to improve accuracy of predicting the ground truth?

Evaluation compares top-$3$ accuracy and mean reciprocal rank (MRR) over samples in a held out test set. Accuracy measures the likelihood that the system scores the ground truth within the top-$3$ candidates against the distractors. MRR compares average ground truth ranking across conditions. A text2text Transformer language model was trained on the OpenSubtitles dataset \cite{lison2018opensubtitles} to predict an output line given a set of input lines \cite{vaswani17attention}. Additional details are in the supplementary material.

The trained Transformer model was used to assign a perplexity score for output line candidates given an input context line. For each unique subtitle file in the validation and test sets, the concatenation of the first $5$ lines serve as input context and line $6$ is the ground truth output to be predicted. Negative candidates are randomly selected from lines in the respective corresponding data segment (i.e. validation or test sets), thus may not be from the same source file as the input context lines.

The perplexity under the trained conversation model serves as the unmodulated probability $q(x_t|x_{1:t-1})$ (Eq. \ref{eqn:tq}) of selection in the prediction task. The input sequence is then passed, line-by-line, through a Newsgroups universe model and a score is assigned to each candidate relative to the change in entropy of the evolving posterior. The $\alpha$ value is modulated over $100$ evenly spaced values between $[-2, 2]$. The accuracy of predicting the ground truth in the top-$3$ candidates and the MRR of the ground truth are computed.

\begin{figure*}[ht]
  \centering
  \includegraphics[width=.9\textwidth]{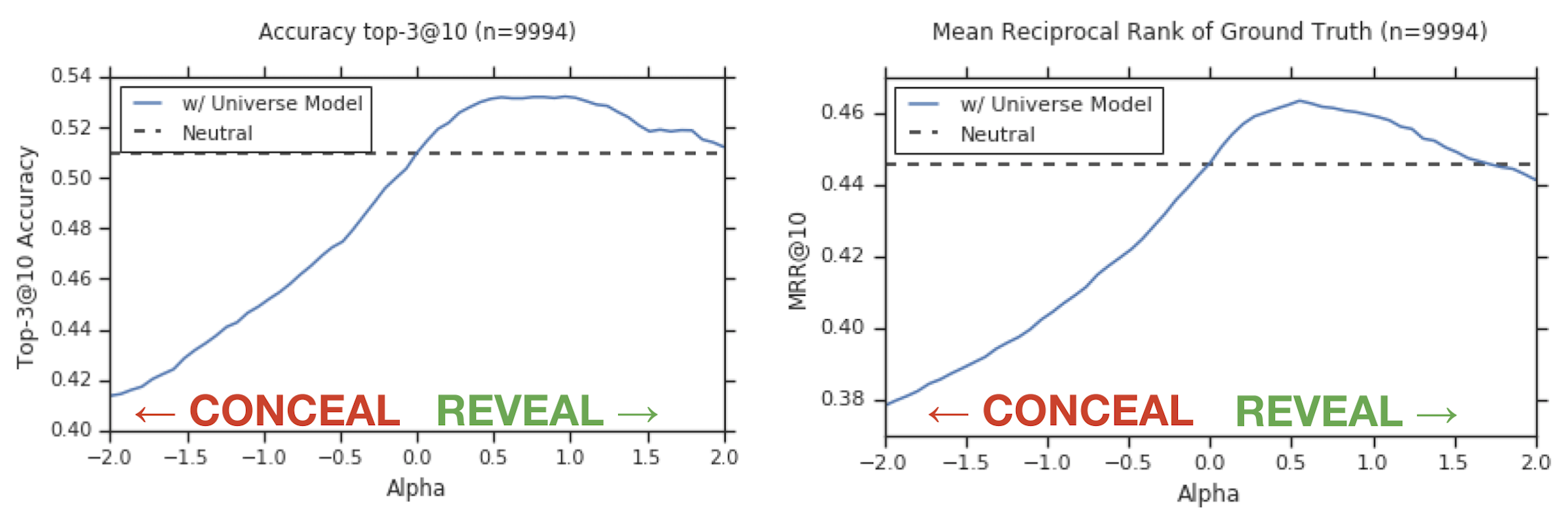}
  \caption[Narrative Arc Boosting Predictive Accuracy]{Information revelation region as $\alpha$ varies for (left) top-$3$ accuracy and (right) MRR in universe model modulated prediction task}
  \label{fig:goldilocks}
\end{figure*}

The results on the validation set are shown in Fig. \ref{fig:goldilocks}. By selecting the correct $\alpha$ value, the likelihood of correctly selecting utterances revealing an incremental amount of information increases significantly. Note the shape of the curve as $\alpha$ changes. As hypothesized, there exists a region, between $0$ and $1$ where the `right' amount of universe information is revealed. This region corresponds to the notion that each line of dialogue will reveal some, but not too much, information about the universe. As $\alpha$ continues to increase, the accuracy decreases below the neutral baseline. The top-$3$ accuracy of prediction increases when the universe model boosts the probabilities of appropriately revealing dialogue. The validation set is used to set the optimal $\alpha$, which is then used to score samples in the test set and results are presented in Table 1. Two additional models are included for comparison. \emph{T2T@1} uses $1$ preceding the ground truth as context. \emph{Unigram} assigns a perplexity to output candidates by building a unigram language model using the $5$ input lines as a corpus. A smoothing factor of $1 \times 10^{-5}$ is used for out-of-vocabulary words. Additionally, a random conversation baseline model is included. For each model tested, information from the universe model significantly improves the predictive accuracy on this task.
\begin{table}[]
\centering
\begin{tabular}{|ll | ll|}
\hline
\textbf{CM}  & \textbf{UM} & \textbf{Top3Acc} & \textbf{MRR} \\
\hline \hline
T2T@5     & NG      & \textbf{0.520} & \textbf{0.456}* \\
T2T@5     & Neutral & 0.507 & 0.444 \\
T2T@1     & NG      & 0.483 & 0.428* \\
T2T@1     & Neutral & 0.469 & 0.412 \\
Unigram   & NG      & 0.366 & 0.337* \\
Unigram   & Neutral & 0.296 & 0.290 \\
\hline
Random    & Neutral & 0.302 & 0.294 \\
\hline
\end{tabular}
\label{table_goldilocks}
\caption{Results for predicting the next line. CM is the conversation model, UM is the universe model, Top3Acc is the accuracy of predicting the ground-truth in the top-$3$ of $10$ candidates, and MRR is the mean reciprocal rank of the ground truth. Unigram CM calculates the perplexity of each candidate given the input lines as training corpus. T2T@N is a Tensor2Tensor Transformer model which uses the previous N lines as an input to predict the output and NG is the Newsgroups universe. A Neutral universe model represents no modulation which is equivalent to $\alpha = 0$. $*$ indicates $p < 0.05$ for a Students' t-test comparing MRR to the Neutral model.}
\end{table}
\subsection{Interactive Collaborative Dialogue} \label{sec:interactive}
Finally, as a practical implementation case-study, we tested how this system performs in collaborative dialogue through interaction with humans. Four expert improvisational theatre performers engaged  with the system in $3$ text-based conversations. Each conversation consisted of $5$ utterance-response pairs for a total of ten utterances (i.e. an average length of a short-duration improvised scene \cite{sawyer2003improvised}). Subjects are native English speakers with 5+ years professional performance experience and are familiar with shared narrative development and collaborative dialogue. Each interacted with revealing, concealing, and neutral models in a randomized order unknown to the them. Transcripts of actor-system dialogues and qualitative feedback are included in supplementary material.

As described in Sec. \ref{sec:conversation_model}, this experiment used the Books conversation model and the DeepMoji universe model. Following the interactions, each performer was asked the following question: ``please rank the conversations from 1 (most engaging) to $3$ (least engaging)''. Engagingness was defined to align with the notions of reaveling and concealing in this work. An agent is engaging for shared scene development if it brings just enough information to add specificity and reduce ambiguity but not limit the conversation.

Three of the four performers ranked the revealing model, $\alpha > 0$, as the most engaging. Those three performers ranked $\alpha = 0$ as being less engaging due to being ``too random''. All subjects ranked $\alpha < 0$ as being least engaging and not bringing enough information to the scene. These results support the hypothesis that $\alpha$ can effectively modulate the engagingness of collaborative dialogue in human-machine interaction.

\section{Related Work} \label{sec:related_work}

Collaborative dialogue between humans and machines has been proposed as a grand challenge in artificial intelligence \cite{mathewson2017improvisedcomedyturing,martin2016improvisational,brown2008auslander}. Previous methods have used hard coded rules, decision trees, and event representations to generate novel narrative chains \cite{Martin2017ImprovisationalSA}.
We used an deep neural network-based generative language model enhanced with universe model information in the context of improvised theatre \cite{mathewson2017improvised}.

While neural response generation systems provide a trainable end-to-end system for language generation, these methods are prone to providing generic, unspecific responses \cite{DBLP:journals/corr/LiGBGD15}. Recent advances have improved generated responses by optimizing sentence encoding and decoding jointly, post-generation candidate re-scoring \cite{bordes2016learning,vinyals2015neural,sordoni2015neural}, reinforcement learning \cite{2016arXiv160601541L}, hierarchical models for distilling extended context \cite{serban2016building}, and auxiliary training objectives, such as maximizing mutual information \cite{DBLP:journals/corr/LiGBGD15}, and personality specificity and consistency \cite{li2016persona,zhang2018personalizing}. In future work, universe models and conversational models could be trained jointly.

Our work is related to the controlled generation of text using disentangled latent representations \cite{hu2017toward,zhou2017emotional,asghar2018affective}. Previous work has used a topic-transition generative adversarial network to enforce smoothness of transition of subsequent utterances \cite{liang2017recurrent}. These methods use neural encoder-decoders and generate responses given an input sequence and a desired target class for the response.

Other, recent work has aimed to improve candidates returned by retrieval-based conversation models \cite{weston2018retrieve}. These methods utilize a conversation model to find similar prototypes using embedding distances and refine prototypes with a sequence-to-sequence model \cite{guu2017generating}. We do not refine candidates from the conversation model, rather we sample and select using a scoring function defined by the revealing and concealing parameter.

Similar to universe models, topic models or lexical fields have been shown capable of tracking general subjects of a text \cite{blei2003latent,geeraerts2010theories}. Dynamic topic models characterize the evolution of topics over a set of documents over time \cite{blei2006dynamic}. 
Our work differs in that we generate dialogue using the evolving probabilistic belief during a single conversation, as opposed to tracking topical shifts over longer timescales. Using a probabilistic classifier for narrative tracking has been explored previously \cite{mohammad2011once,reagan2016emotional}. These works used sentiment classifiers to track emotion and plots arcs through narratives. We extend these works by using probabilistic universe models collaborative dialogue generation.

\section{Discussion and Conclusion}
\label{sec:conclusion}
While innovations have improved the linguistic quality, semantic alignment, and consistency of utterances generated by neural models, generated conversations still lack interestingness and engagingness. Our work generates engaging dialogue by shaping the underlying narrative arc as opposed to improving the training of generative language models. The methods presented are agnostic to both the universe and the conversational model used.
Using rules from improvised theatre, we quantitatively define the evolution of interesting and engaging dialogue.

In this work we focus on genre, emoji, and topic-based universe models. Other universe models to be explored involve causality of events, directions of relationships, or audience reaction prediction \cite{riedl2006linear,knight2011savvy,trabasso1985causal,cook1928plotto,eger2015plotter}.
While this work explores the interaction between a base conversation model and a universe model, this method could be compatible with image or video generation.

The main contribution of this work is the computational formalization of the narrative arc, an information-theoretic framework for collaborative dialogue interaction. The framework fills a gap in previous research by connecting the utterance-level improvements of language models with the conversation-level improvements of universe tracking. This is done by sampling candidates from a conversational model using a universe model and the narrative arc.
We illustrate narrative arcs underlying popular dialogues and show how universe models can be combined with conversation models to generate interesting dialogue. We present empirical results showing how the narrative arc can improve accuracy on a next line prediction task. Finally, we present an expert user-study to validate our model.

\bibliography{narrative-arc-arxiv-2019}

\begin{thebibliography}{56}
\providecommand{\natexlab}[1]{#1}
\providecommand{\url}[1]{\texttt{#1}}
\expandafter\ifx\csname urlstyle\endcsname\relax
  \providecommand{\doi}[1]{doi: #1}\else
  \providecommand{\doi}{doi: \begingroup \urlstyle{rm}\Url}\fi

\bibitem[Asghar et~al.(2018)Asghar, Poupart, Hoey, Jiang, and
  Mou]{asghar2018affective}
Asghar, N., Poupart, P., Hoey, J., Jiang, X., and Mou, L.
\newblock Affective neural response generation.
\newblock In \emph{Eur. Conference on Information Retrieval}, pp.\  154--166.
  Springer, 2018.

\bibitem[Bickmore \& Giorgino(2006)Bickmore and Giorgino]{BICKMORE2006556}
Bickmore, T. and Giorgino, T.
\newblock Health dialog systems for patients and consumers.
\newblock \emph{J. Biomedical Informatics}, 39\penalty0 (5):\penalty0 556 --
  571, 2006.
\newblock ISSN 1532-0464.
\newblock Dialog Systems for Health Communications.

\bibitem[Bishop(2006)]{bishop06pattern}
Bishop, C.~M.
\newblock \emph{Pattern Recognition and Machine Learning}.
\newblock Springer, New York, 2006.

\bibitem[Bizzocchi(2007)]{bizzocchi2007games}
Bizzocchi, J.
\newblock Games and narrative: An analytical framework.
\newblock \emph{Loading-The Journal of the Canadian Games Studies Association},
  1\penalty0 (1):\penalty0 5--10, 2007.

\bibitem[Blei \& Lafferty(2006)Blei and Lafferty]{blei2006dynamic}
Blei, D.~M. and Lafferty, J.~D.
\newblock Dynamic topic models.
\newblock In \emph{Proc. of the 23rd Intl. Conf. on Machine learning}, pp.\
  113--120. ACM, 2006.

\bibitem[Blei et~al.(2003)Blei, Ng, and Jordan]{blei2003latent}
Blei, D.~M., Ng, A.~Y., and Jordan, M.~I.
\newblock Latent dirichlet allocation.
\newblock \emph{J. of Machine Learning Research}, 3\penalty0 (Jan):\penalty0
  993--1022, 2003.

\bibitem[Bordes et~al.(2016)Bordes, Boureau, and Weston]{bordes2016learning}
Bordes, A., Boureau, Y.-L., and Weston, J.
\newblock Learning end-to-end goal-oriented dialog.
\newblock \emph{arXiv preprint arXiv:1605.07683}, 2016.

\bibitem[Brown(2008)]{brown2008auslander}
Brown, K.
\newblock The auslander test: or,‘of bots and humans’.
\newblock \emph{Intl. J. Performance Arts and Digital Media}, 4\penalty0
  (2-3):\penalty0 181--188, 2008.

\bibitem[Cer et~al.(2018)Cer, Yang, Kong, Hua, Limtiaco, John, Constant,
  Guajardo-Cespedes, Yuan, Tar, et~al.]{cer2018universal}
Cer, D., Yang, Y., Kong, S.-y., Hua, N., Limtiaco, N., John, R.~S., Constant,
  N., Guajardo-Cespedes, M., Yuan, S., Tar, C., et~al.
\newblock Universal sentence encoder.
\newblock \emph{arXiv preprint arXiv:1803.11175}, 2018.

\bibitem[Cesa-Bianchi \& Lugosi(2006)Cesa-Bianchi and
  Lugosi]{cesabianchi06prediction}
Cesa-Bianchi, N. and Lugosi, G.
\newblock \emph{{Prediction, Learning, and Games}}.
\newblock Cambridge University Press, New York, NY, USA, 2006.

\bibitem[Cook(1928)]{cook1928plotto}
Cook, W.~W.
\newblock \emph{Plotto: a new method of plot suggestion for writers of creative
  fiction}.
\newblock Ellis., 1928.

\bibitem[Eger et~al.(2015)Eger, Potts, Barot, and Young]{eger2015plotter}
Eger, M., Potts, C.~M., Barot, C., and Young, R.~M.
\newblock Plotter: operationalizing the master book of all plots.
\newblock \emph{Proceedings of the Intelligent Narrative Technologies and
  Social Believability in Games}, pp.\  30--33, 2015.

\bibitem[{Felbo} et~al.(2017){Felbo}, {Mislove}, {S{\o}gaard}, {Rahwan}, and
  {Lehmann}]{2017arXiv170800524F}
{Felbo}, B., {Mislove}, A., {S{\o}gaard}, A., {Rahwan}, I., and {Lehmann}, S.
\newblock {Using millions of emoji occurrences to learn any-domain
  representations for detecting sentiment, emotion and sarcasm}.
\newblock \emph{ArXiv e-prints}, August 2017.

\bibitem[Fitzpatrick et~al.(2017)Fitzpatrick, Darcy, and
  Vierhile]{fitzpatrick17delivering}
Fitzpatrick, K.~K., Darcy, A., and Vierhile, M.
\newblock Delivering cognitive behavior therapy to young adults with symptoms
  of depression and anxiety using a fully automated conversational agent
  (woebot): A randomized controlled trial.
\newblock \emph{JMIR Ment Health}, 4\penalty0 (2):\penalty0 e19, Jun 2017.

\bibitem[Geeraerts(2010)]{geeraerts2010theories}
Geeraerts, D.
\newblock \emph{Theories of lexical semantics}.
\newblock Oxford University Press, 2010.

\bibitem[Graves(2013)]{graves2013generating}
Graves, A.
\newblock Generating sequences with recurrent neural networks.
\newblock \emph{arXiv preprint arXiv:1308.0850}, 2013.

\bibitem[Guu et~al.(2017)Guu, Hashimoto, Oren, and Liang]{guu2017generating}
Guu, K., Hashimoto, T.~B., Oren, Y., and Liang, P.
\newblock Generating sentences by editing prototypes.
\newblock \emph{arXiv preprint arXiv:1709.08878}, 2017.

\bibitem[Hoang(2018)]{hoang2018predicting}
Hoang, Q.
\newblock Predicting movie genres based on plot summaries.
\newblock \emph{arXiv preprint arXiv:1801.04813}, 2018.

\bibitem[Hu et~al.(2017)Hu, Yang, Liang, Salakhutdinov, and Xing]{hu2017toward}
Hu, Z., Yang, Z., Liang, X., Salakhutdinov, R., and Xing, E.~P.
\newblock Toward controlled generation of text.
\newblock In \emph{Intl. Conf. on Machine Learning}, pp.\  1587--1596, 2017.

\bibitem[Joachims(1996)]{joachims1996probabilistic}
Joachims, T.
\newblock A probabilistic analysis of the rocchio algorithm with tfidf for text
  categorization.
\newblock Technical report, CMU, 1996.

\bibitem[Johnstone(1979)]{johnstone1979impro}
Johnstone, K.
\newblock \emph{Impro: Improvisation and the Theatre}.
\newblock Faber and Faber Ltd., 1979.

\bibitem[Knight et~al.(2011)Knight, Satkin, Ramakrishna, and
  Divvala]{knight2011savvy}
Knight, H., Satkin, S., Ramakrishna, V., and Divvala, S.
\newblock A savvy robot standup comic: Online learning through audience
  tracking.
\newblock In \emph{Workshop paper (TEI'10)}, 2011.

\bibitem[Kripke(1963)]{kripke1963semantical}
Kripke, S.~A.
\newblock Semantical analysis of modal logic i normal modal propositional
  calculi.
\newblock \emph{Mathematical Logic Quarterly}, 9\penalty0 (5-6):\penalty0
  67--96, 1963.

\bibitem[Li et~al.(2015)Li, Galley, Brockett, Gao, and
  Dolan]{DBLP:journals/corr/LiGBGD15}
Li, J., Galley, M., Brockett, C., Gao, J., and Dolan, B.
\newblock A diversity-promoting objective function for neural conversation
  models.
\newblock \emph{CoRR}, abs/1510.03055, 2015.

\bibitem[Li et~al.(2016)Li, Galley, Brockett, Spithourakis, Gao, and
  Dolan]{li2016persona}
Li, J., Galley, M., Brockett, C., Spithourakis, G.~P., Gao, J., and Dolan, B.
\newblock A persona-based neural conversation model.
\newblock \emph{arXiv preprint arXiv:1603.06155}, 2016.

\bibitem[{Li} et~al.(2016){Li}, {Monroe}, {Ritter}, {Galley}, {Gao}, and
  {Jurafsky}]{2016arXiv160601541L}
{Li}, J., {Monroe}, W., {Ritter}, A., {Galley}, M., {Gao}, J., and {Jurafsky},
  D.
\newblock {Deep Reinforcement Learning for Dialogue Generation}.
\newblock \emph{ArXiv e-prints}, June 2016.

\bibitem[Liang et~al.(2017)Liang, Hu, Zhang, Gan, and Xing]{liang2017recurrent}
Liang, X., Hu, Z., Zhang, H., Gan, C., and Xing, E.~P.
\newblock Recurrent topic-transition gan for visual paragraph generation.
\newblock \emph{CoRR, abs/1703.07022}, 2, 2017.

\bibitem[Lison et~al.(2018)Lison, Tiedemann, and
  Kouylekov]{lison2018opensubtitles}
Lison, P., Tiedemann, J., and Kouylekov, M.
\newblock Opensubtitles 2018: Statistical rescoring of sentence alignments in
  large, noisy parallel corpora.
\newblock In \emph{Proc. 11e LREC 2018, Miyazaki, Japan.}, 2018.

\bibitem[Liu et~al.(2018)Liu, Saleh, Pot, Goodrich, Sepassi, Kaiser, and
  Shazeer]{liu2018generating}
Liu, P.~J., Saleh, M., Pot, E., Goodrich, B., Sepassi, R., Kaiser, L., and
  Shazeer, N.
\newblock Generating wikipedia by summarizing long sequences.
\newblock \emph{arXiv preprint arXiv:1801.10198}, 2018.

\bibitem[Martin et~al.(2016)Martin, Harrison, and
  Riedl]{martin2016improvisational}
Martin, L.~J., Harrison, B., and Riedl, M.~O.
\newblock Improvisational computational storytelling in open worlds.
\newblock In \emph{Intl. Conf. on Interactive Digital Storytelling}, pp.\
  73--84. Springer, 2016.

\bibitem[Martin et~al.(2017)Martin, Ammanabrolu, Wang, Singh, Harrison,
  Dhuliawala, Tambwekar, Mehta, Arora, Dass, Purdy, and
  Riedl]{Martin2017ImprovisationalSA}
Martin, L.~J., Ammanabrolu, P., Wang, X., Singh, S., Harrison, B., Dhuliawala,
  M., Tambwekar, P., Mehta, A.~A., Arora, R., Dass, N., Purdy, C., and Riedl,
  M.~O.
\newblock Improvisational storytelling agents.
\newblock In \emph{Proc. of the Machine Learning for Creativity and Design
  Workshop at NIPS 2017}, 2017.

\bibitem[Mathewson \& Mirowski(2017{\natexlab{a}})Mathewson and
  Mirowski]{mathewson2017improvised}
Mathewson, K.~W. and Mirowski, P.
\newblock Improvised theatre alongside artificial intelligences.
\newblock In \emph{AAAI AIIDE}, 2017{\natexlab{a}}.

\bibitem[Mathewson \& Mirowski(2017{\natexlab{b}})Mathewson and
  Mirowski]{mathewson2017improvisedcomedyturing}
Mathewson, K.~W. and Mirowski, P.
\newblock Improvised comedy as a turing test.
\newblock In \emph{Proc. of the Machine Learning for Creativity and Design
  Workshop at NIPS 2017}, 2017{\natexlab{b}}.

\bibitem[McLeod(2000)]{mcleod2000use}
McLeod, C.
\newblock \emph{The use of Keith Johnstone's concepts of Circles of Expectation
  and Vogler's generic story structure as directorial aids in comedy theatre}.
\newblock PhD thesis, University of Cape Town, 2000.

\bibitem[Miner et~al.(2017)Miner, Milstein, and
  Hancock]{doi:10.1001/jama.2017.14151}
Miner, A., Milstein, A., and Hancock, J.
\newblock Talking to machines about personal mental health problems.
\newblock \emph{JAMA}, 318\penalty0 (13):\penalty0 1217--1218, 2017.

\bibitem[Mohammad(2011)]{mohammad2011once}
Mohammad, S.
\newblock From once upon a time to happily ever after: Tracking emotions in
  novels and fairy tales.
\newblock In \emph{Proc. of the 5th ACL-HLT Workshop on Language Technology for
  Cultural Heritage, Social Sciences, and Humanities}, pp.\  105--114. ACL,
  2011.

\bibitem[Murphy(2012)]{murphy2012prob}
Murphy, K.~P.
\newblock \emph{Machine Learning: A Probabilistic Perspective}.
\newblock The MIT Press, 2012.

\bibitem[Nikitina et~al.(2018)Nikitina, Callaioli, and Baez]{nikitina2018smart}
Nikitina, S., Callaioli, S., and Baez, M.
\newblock Smart conversational agents for reminiscence.
\newblock \emph{arXiv preprint arXiv:1804.06550}, 2018.

\bibitem[Raby(2010)]{raby2010improvisation}
Raby, G.
\newblock Improvisation and devising: The circle of expectation, the invisible
  hand, and rsvp.
\newblock \emph{Canadian Theatre Review}, 143\penalty0 (1):\penalty0 94--97,
  2010.

\bibitem[Reagan et~al.(2016)Reagan, Mitchell, Kiley, Danforth, and
  Dodds]{reagan2016emotional}
Reagan, A.~J., Mitchell, L., Kiley, D., Danforth, C.~M., and Dodds, P.~S.
\newblock The emotional arcs of stories are dominated by six basic shapes.
\newblock \emph{EPJ Data Science}, 5\penalty0 (1):\penalty0 31, 2016.

\bibitem[Riedl \& Young(2006)Riedl and Young]{riedl2006linear}
Riedl, M.~O. and Young, R.~M.
\newblock From linear story generation to branching story graphs.
\newblock \emph{IEEE Computer Graphics and Applications}, 26\penalty0
  (3):\penalty0 23--31, 2006.

\bibitem[Salton \& Buckley(1988)Salton and Buckley]{salton1988term}
Salton, G. and Buckley, C.
\newblock Term-weighting approaches in automatic text retrieval.
\newblock \emph{Information processing \& management}, 24\penalty0
  (5):\penalty0 513--523, 1988.

\bibitem[Sawyer(2003)]{sawyer2003improvised}
Sawyer, R.~K.
\newblock \emph{Improvised dialogues: Emergence and creativity in
  conversation}.
\newblock Greenwood Publishing Group, 2003.

\bibitem[Sennrich et~al.(2015)Sennrich, Haddow, and Birch]{sennrich2015neural}
Sennrich, R., Haddow, B., and Birch, A.
\newblock Neural machine translation of rare words with subword units.
\newblock \emph{arXiv preprint arXiv:1508.07909}, 2015.

\bibitem[Serban et~al.(2016)Serban, Sordoni, Bengio, Courville, and
  Pineau]{serban2016building}
Serban, I.~V., Sordoni, A., Bengio, Y., Courville, A., and Pineau, J.
\newblock Building end-to-end dialogue systems using generative hierarchical
  neural network models.
\newblock In \emph{Proc. AAAI}, AAAI'16, pp.\  3776--3783. AAAI Press, 2016.

\bibitem[Shannon(1951)]{shannon1951prediction}
Shannon, C.~E.
\newblock Prediction and entropy of printed english.
\newblock \emph{Bell Labs Tech. J.}, 30\penalty0 (1):\penalty0 50--64, 1951.

\bibitem[Sordoni et~al.(2015)Sordoni, Galley, Auli, Brockett, Ji, Mitchell,
  Nie, Gao, and Dolan]{sordoni2015neural}
Sordoni, A., Galley, M., Auli, M., Brockett, C., Ji, Y., Mitchell, M., Nie,
  J.-Y., Gao, J., and Dolan, B.
\newblock A neural network approach to context-sensitive generation of
  conversational responses.
\newblock \emph{arXiv preprint arXiv:1506.06714}, 2015.

\bibitem[Swain(2000)]{swain2000output}
Swain, M.
\newblock The output hypothesis and beyond: Mediating acquisition through
  collaborative dialogue.
\newblock \emph{Sociocultural Theory and Second Language Learning},
  97:\penalty0 114, 2000.

\bibitem[Trabasso \& Sperry(1985)Trabasso and Sperry]{trabasso1985causal}
Trabasso, T. and Sperry, L.~L.
\newblock Causal relatedness and importance of story events.
\newblock \emph{Journal of Memory and language}, 24\penalty0 (5):\penalty0 595,
  1985.

\bibitem[Vaswani et~al.(2017)Vaswani, Shazeer, Parmar, Uszkoreit, Jones, Gomez,
  Kaiser, and Polosukhin]{vaswani17attention}
Vaswani, A., Shazeer, N., Parmar, N., Uszkoreit, J., Jones, L., Gomez, A.~N.,
  Kaiser, L., and Polosukhin, I.
\newblock Attention is all you need.
\newblock \emph{CoRR}, abs/1706.03762, 2017.

\bibitem[{Venkatesh} et~al.(2018){Venkatesh}, {Khatri}, {Ram}, {Guo},
  {Gabriel}, {Nagar}, {Prasad}, {Cheng}, {Hedayatnia}, {Metallinou}, {Goel},
  {Yang}, and {Raju}]{venkatesh2018evaluating}
{Venkatesh}, A., {Khatri}, C., {Ram}, A., {Guo}, F., {Gabriel}, R., {Nagar},
  A., {Prasad}, R., {Cheng}, M., {Hedayatnia}, B., {Metallinou}, A., {Goel},
  R., {Yang}, S., and {Raju}, A.
\newblock On evaluating and comparing conversational agents.
\newblock \emph{arXiv preprint arXiv:1801.03625}, 2018.

\bibitem[Vinyals \& Le(2015)Vinyals and Le]{vinyals2015neural}
Vinyals, O. and Le, Q.
\newblock A neural conversational model.
\newblock \emph{arXiv preprint arXiv:1506.05869}, 2015.

\bibitem[Weizenbaum(1966)]{weizenbaum1966eliza}
Weizenbaum, J.
\newblock Eliza: a computer program for the study of natural language
  communication between man and machine.
\newblock \emph{Comm. of the ACM}, 9\penalty0 (1):\penalty0 36--45, 1966.

\bibitem[Weston et~al.(2018)Weston, Dinan, and Miller]{weston2018retrieve}
Weston, J., Dinan, E., and Miller, A.~H.
\newblock Retrieve and refine: Improved sequence generation models for
  dialogue.
\newblock \emph{arXiv preprint arXiv:1808.04776}, 2018.

\bibitem[Zhang et~al.(2018)Zhang, Dinan, Urbanek, Szlam, Kiela, and
  Weston]{zhang2018personalizing}
Zhang, S., Dinan, E., Urbanek, J., Szlam, A., Kiela, D., and Weston, J.
\newblock Personalizing dialogue agents: {I} have a dog, do you have pets too?
\newblock \emph{CoRR}, abs/1801.07243, 2018.

\bibitem[Zhou et~al.(2017)Zhou, Huang, Zhang, Zhu, and Liu]{zhou2017emotional}
Zhou, H., Huang, M., Zhang, T., Zhu, X., and Liu, B.
\newblock Emotional chatting machine: emotional conversation generation with
  internal and external memory.
\newblock \emph{arXiv preprint arXiv:1704.01074}, 2017.

\end{thebibliography}
\bibliographystyle{icml2019}

\clearpage
\onecolumn

\section{Supplementary Material}

\begin{appendices}

\section{Data Processing Details}

OpenSubtitles were used as conversation model data.\footnote{\url{http://opus.nlpl.eu/OpenSubtitles.php}}. The dataset was preprocessed by removing duplicate movie subtitle files, lines under 10 characters and duplicate lines, resulting in 68,719,885 unique lines.
The text2text Transformer model from Google's open source implementation was used for training.\footnote{\url{https://github.com/tensorflow/tensor2tensor}} OpenSubtitles were used as training data.\footnote{\url{http://opus.nlpl.eu/OpenSubtitles.php}} As several files in the dataset cover the same uniquely identified movie or television show duplicates were removed by keeping only the subtitle file with the most lines for each unique ID.  The dataset was split into into $119,949$ training IDs, $10,000$ validation IDs, and $10,000$ testing IDs. The data was prepared for training by removing empty lines, duplicate lines, and substituting non-Unicode characters. A vocabulary was built using the training set. After cleaning, subtitle files with less than $6$ lines were excluded. Training data was formatted into input/response pairs. The training data was split into $17,136,469$ training examples and $1,437,473$ evaluation examples. Validation and testing subtitle sets were held out to measure task accuracy on unseen data. The hyperparameters of the Transformer model were set as follows: hidden size of $1024$, filter size of $4096$, batch size of $2048$, $16$ heads, and a dropout factor of $0.3$ was used for regularization. The model was trained for $861,000$ steps, to convergence, with final negative log-perplexity of $-2.976$ on evaluation set.

\section{Rejection Sampling} \label{appendix:rejection_sampling}

Rejection sampling is a means of online sampling that allows for sampling from an unknown distribution. Suppose we are given an unnormalized distribution $\tq$ over $\cX$ which we can query (i.e. $\tq(x)$) but not necessarily integrate over. Let $q$ be a proper distribution over $\cX$ such that $\tq$ is dominated by $qM$, $M \in \bR$:
\begin{equation}
  \label{eqn:qm_domination}
  \tq(x) \le M q(x), \qquad \forall \, x \in \cX .
\end{equation}
The \emph{Rejection sampling} algorithm to obtain a sample from an unnormalized distribution $\tq$ using samples from a proper distribution $q$ proceeds as follows:

\begin{enumerate}
\setlength\itemsep{-0.2em}
\item Sample $x \sim q$ and sample $r \sim \text{Uniform}(0, 1)$,
\item If $r \le \tq(x) / Mq(x)$, accept $x$ as a sample drawn from $\tq$, otherwise reject the sample and go to 1.
\end{enumerate}
This algorithm will take an average of $M$ iterations to obtain a sample. Let $X$ be the random element returned by this procedure.
\begin{prop}
Rejection sampling samples $x \in \cX$ with probability
  \begin{equation*}
    \Pr\{ X = x \} = \frac{\tq(x)}{\sum_{x' \in \cX} \tq(x') } .
  \end{equation*}
\end{prop}
\begin{proof}
Write $x_1, x_2, \dots$ for the sequence of sampled symbols, and write $A_i$ to denote the event
that the $i^{th}$ symbol is accepted (at which point we stop the process).
Then
\begin{align*}
\Pr \{ X = x \} & = \Pr \{ A_1, x_1 = x \} + (1 - \Pr \{ A_1 \}) \Pr \{ A_2, x_2 = x \} + \dots \\
&= \sum_{t=1}^\infty \Pr \{ A_t, x_t = x \} \prod_{s=1}^{t-1} (1 - \Pr \{ A_s \}) .
\end{align*}
Now, this process is memoryless (in a sense we should make a little more formal) and
\begin{equation*}
\Pr \{ A_s \} = \Pr \{ A_1 \} \qquad \Pr \{A_t, x_t = x \} = \Pr \{ A_1, x_1 = x \} .
\end{equation*}
Since $1 - \Pr \{ A_1 \} < 1$, the geometric sum converges and
\begin{equation*}
\Pr \{ X = x \} = \frac{ \Pr \{A_1, x_1 = x\}}{ \Pr \{ A_1 \} },
\end{equation*}
which is the conditional probability of $x_1 = x$ given $A_1$. Now
\begin{equation*}
\Pr \{A_1, x_1 = x\} = q(x) \Pr \left \{r \le \frac{\tq(x)}{M q(x)} \right \} = q(x) \frac{\tq(x)}{M q(x)} = \frac{\tq(x)}{M},
\end{equation*}
where we used \eqnref{qm_domination} to guarantee that $\frac{\tq(x)}{M q(x)} < 1$. But then
\begin{equation*}
\Pr \{A_1\} = \sum_{x' \in \cX} \Pr \{A_1, x_1 = x' \} = \frac{1}{M} \sum_{x' \in \cX} \tq(x') .
\end{equation*}
We conclude that
\begin{equation*}
\Pr \{X = x \} = \frac{\Pr \{ A_1, x_1 = x\} } {\Pr \{ A_1 \}} = \frac{\tfrac{1}{M} \tq(x)}{\tfrac{1}{M} \sum_{x'} \tq(x')} = \frac{\tq(x)}{\sum_{x'} \tq(x')},
\end{equation*}
as desired.
\end{proof}

\section{Additional Narrative Arcs Underlying Existing Dialogues}
\label{appendix:underlying}

We present an additional set of narrative arcs underlying 20-utterance scenes in Figure 2 and discussed in Section 3.1. In the paper body we report results on the Romeo and Juliet scene.

\begin{enumerate}
\item ``Argument Clinic'' Monty Python’s Flying Circus, Episode 29 (1972). First 20 lines.
\item ``Blade Runner'' Hampton Fancher and David People. Adapted from a novel Philip K. Dick (1982). Last 20 lines of director's cut movie script.
\item ``The Great Dictator'' Charlie Chaplin. Los Angeles, CA: United Artists (1940). First 20 lines of Charlie Chaplin's speech.
\item ``Dead Parrot Sketch'' Monty Python’s Flying Circus, Episode 8 (1969). First 20 lines.
\item ``Romeo and Juliet'' William Shakespeare. Vol. 1. JB Lippincott, (1871). Act 1, Scene 1, first 20 lines.

\end{enumerate}
\begin{figure*}[ht]
  \centering
    \includegraphics[width=\textwidth]{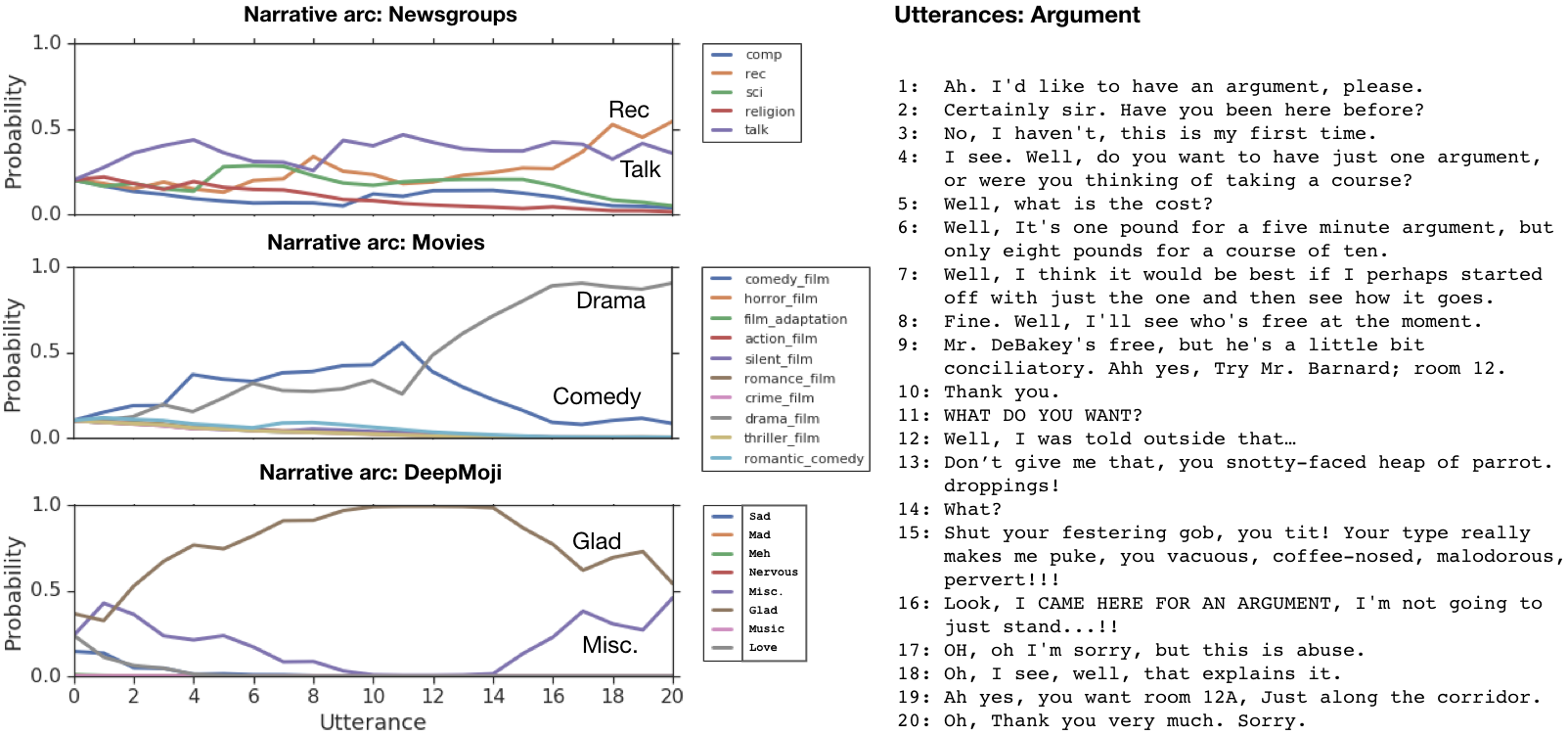}
  \caption{Narrative Arc of ``Argument Clinic''}
\end{figure*}
\begin{figure*}[h]
  \centering
    \includegraphics[width=\textwidth]{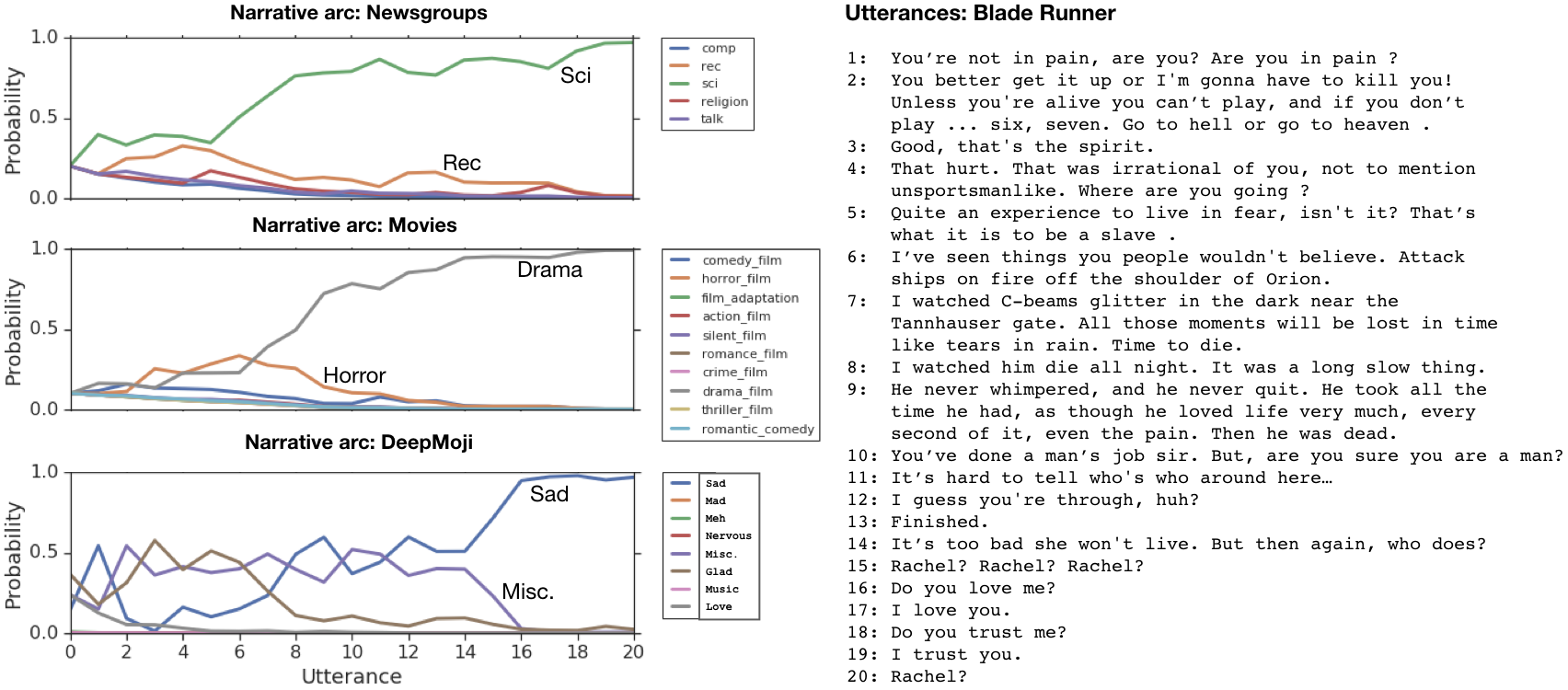}
  \caption{Narrative Arc of ``Blade Runner''}
\end{figure*}
\begin{figure*}[ht]
  \centering
    \includegraphics[width=\textwidth]{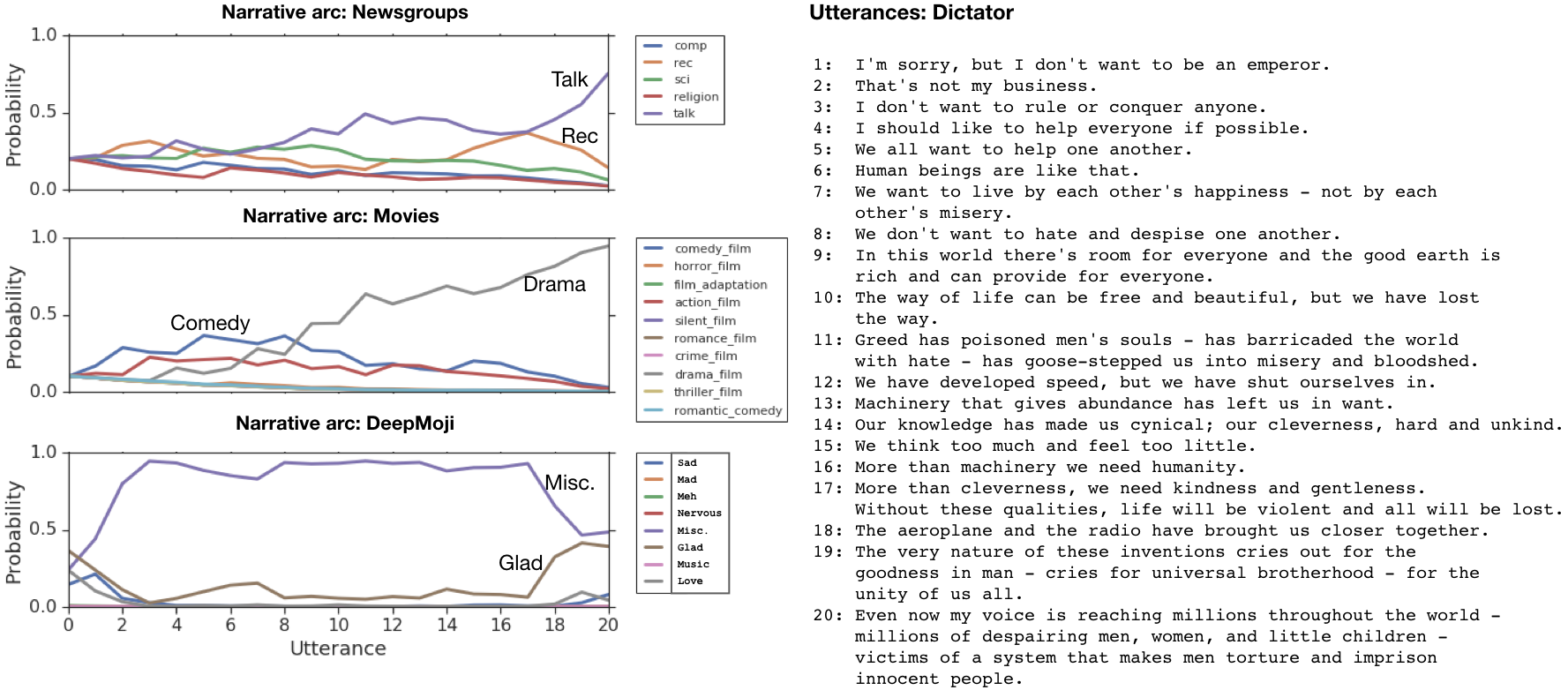}
  \caption{Narrative Arc of ``The Great Dictator''}
\end{figure*}
\begin{figure*}[h]
  \centering
    \includegraphics[width=\textwidth]{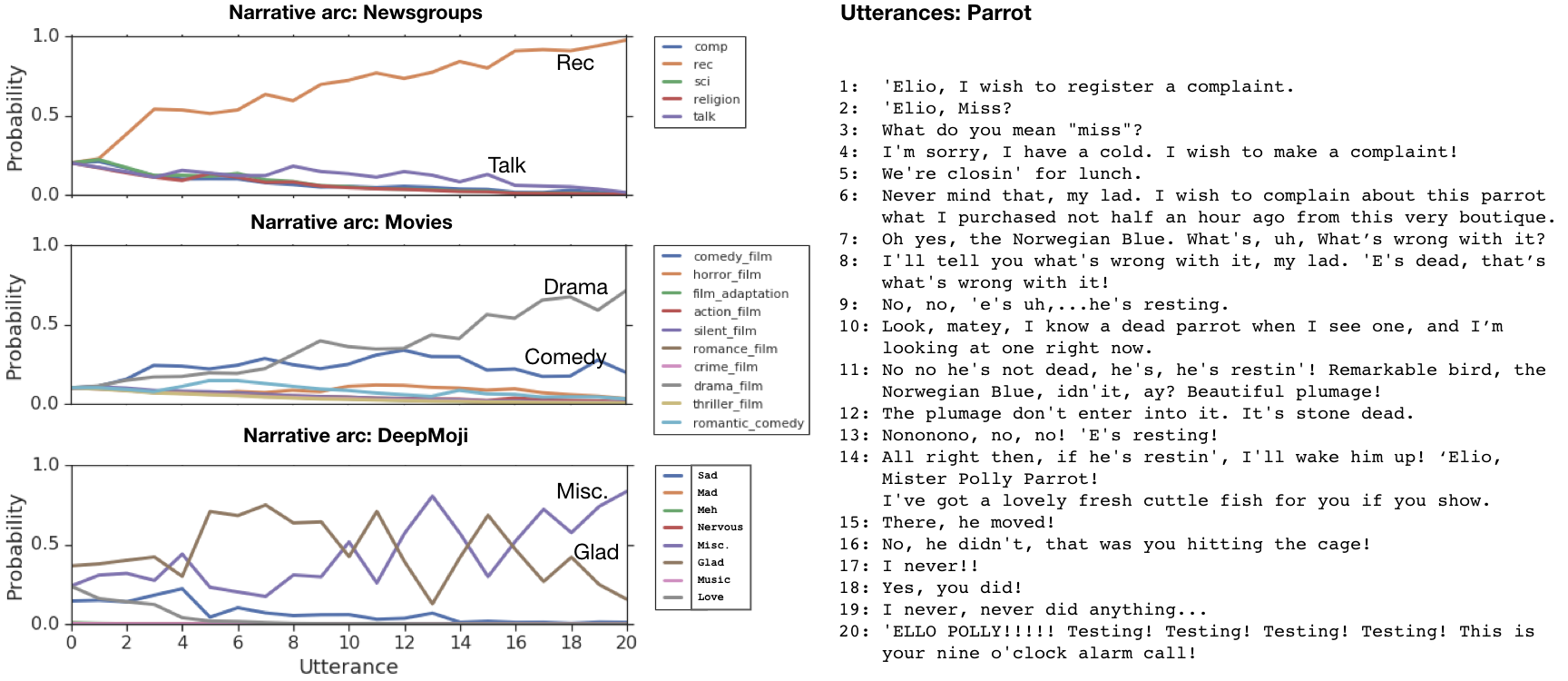}
  \caption{Narrative Arc of ``Dead Parrot Sketch''}
\end{figure*}

\clearpage

\section{Expert Improvisor Conversation Transcripts}
\label{appendix:experttranscripts}

\subsection{Subject1, Mode: Conceal}
\begin{dialogue}
  {
    \speak{Human}Good morning dad, is breakfast ready?
    \speak{System}Breakfast is ready.
    \speak{Human}Great! What are we having?
    \speak{System}We are having Lamb Chops and peas for dinner.
    \speak{Human}Yummmm. But what are we having for breakfast?
    \speak{System}some breakfast.
    \speak{Human}Ok, thanks dad. I have a big ball game later, I would love if you and mom came to watch.
    \speak{System}Likewise, Dad will fire up the chainsaw and be ready to cut through the big limbs at the joints as you work.
    \speak{Human}So I guess you're not coming to the game. You work too much to be good father.
    \speak{System}I'm doing the best I can as a father, and I hope that's good enough.
    \speak{Human}It's not. I'm moving out unless you change.
    \speak{System}If you will leave me to it, I will pack up my things and move out as quickly as I can so as not to inconvenience you.
  }
\end{dialogue}

\subsection{Subject1, Mode: Neutral}
\begin{dialogue}
  {
    \speak{Human}Rosanne, thanks for meeting me under the oak tree
    \speak{System}Keryn would also like to thank her cousins, Dana, Jason, Staci, and Ryan, and her aunt Marge Naegele for continually supporting her decisions even when they thought she was crazy.
    \speak{Human}It means the world that you all came to meet here today as we prepare for another one of my crazy ideas
    \speak{System}We shall all be here at four o’clock.
    \speak{Human}Who else is coming?
    \speak{System}A lot of people are coming.
    \speak{Human}How many?
    \speak{System}How many and where?
    \speak{Human}We need at least 200 here at the oak tree
    \speak{System}Gonna need at least three or four hundred.
    \speak{Human}You're right. How fast can they get here?
    \speak{System}however fast they are travelling.
    \speak{Human}Well I can't argue there. Prepare the tamborines. We all need tamborines for my crazy idea.
    \speak{System}But we're depositing the money—you'll need a chequebook.
  }
\end{dialogue}

\subsection{Subject1, Mode: Reveal}
\begin{dialogue}
  {
    \speak{Human}Welcome to tim hortons, what can I get you?
    \speak{System}I'll give you doughnuts and coffee with milk.
    \speak{Human}Great work Jason, you're doing great on your first day on the job
    \speak{System}First day Welcome the employee!
    \speak{Human}That's right Jason. You're very welcome here. If you do a good job, you'll be a manager in no time.
    \speak{System}Being responsible and knowledgeable is possible in a participative environment if the manager is a member of the overall team.
    \speak{Human}Wow. Jason you continue to surprise me. Where did you learn that?
    \speak{System}``The next time I go up I'll get the information for you,'' Jason told him.
    \speak{Human}Jason? Are you ok?
    \speak{System}JASON: I’m glad you’re feeling better.
    \speak{Human}You're starting to scare me
    \speak{System}And I frighten you.
  }
\end{dialogue}

\subsection{Subject2, Mode: Conceal}
\begin{dialogue}
  {
    \speak{Human}Stacey! How are you? I'm excited that we're getting our nails done for Sams wedding tomorrow!
    \speak{System}I have to tell you I’m so excited for you!
    \speak{Human}Thank you! Maid of Honour is a real honour! Did you remember to get them a gift?
    \speak{System}I am so, so glad you're my maid of honor.
    \speak{Human}And?
    \speak{System}About what?
    \speak{Human}Did you remember to get your parents a gift? They are hosting your wedding after all.
    \speak{System}Otherwise, plan to give them their gifts at the rehearsal dinner.
    \speak{Human}Oh that's a good idea! Hopefully they won't need it earlier.
    \speak{System}They also need it in time so they can get it in the fridge to chill.
  }
\end{dialogue}

\subsection{Subject2, Mode: Neutral}
\begin{dialogue}
  {
    \speak{Human}Good Morning! I was wondering if you had any butcher knives for sale?
    \speak{System}I had an extra one, roast carving size, sent me by the then president of Macy’s last summer, but when Mr. Case sent me two of his sets I gave this odd knife to my part-time cook.
    \speak{Human}Wow! May I see it? I'm getting a... gift... for my boyfriend.
    \speak{System}I sent it to you because you are now a man and it's time for you to look at the future and fulfill the promise I gave to you in the desert—that I'd see you married."
    \speak{Human}Why would I want to make you happy? We haven't spoken in years!
    \speak{System}Because I saw how kind your heart was?
    \speak{Human}I have missed you. I don't know if I have what it takes. Am I ready to commit? Am I ready to take care of someone other than myself?
    \speak{System}I am ready, willing, and worthy to speak with you right here and right now.
    \speak{Human}Will you teach me the butcher ways. So that I may follow in your steps?
    \speak{System}I do it the easy way: I ask the butcher to score it; his knives are better than mine, for a start.
  }
\end{dialogue}

\subsection{Subject2, Mode: Reveal}
\begin{dialogue}
  {
    \speak{Human}Crikey Mate! Did you put the shrimp on the barby yet?
    \speak{System}We started the shrimp over a hot fire, then finished them over low heat to allow them to cook through gently.
    \speak{Human}Well that sounds delicious! We need to impress the Boss tonight! Did Karen make her cheese dip too? I can't get enough of that!
    \speak{System}She makes some and we sit at the table again.
    \speak{Human}Sounds great Greg! Listen, I really appreciate you hosting this party, and I hope you're not too upset that I got the promotion over you.
    \speak{System}(11) If Greg came to the party, all hell would break loose.
    \speak{Human}Yeah I can't stand that guy. Why have two Gregs in the office! You are all we need, am I right?
    \speak{System}He's replacing Laura as our new office (assistant).
    \speak{Human}Sounds like a plan. Well thanks again. Maybe next month you'll get the promotion!
    \speak{System}And you may depend on me to know my place if you get promoted.
  }
\end{dialogue}

\subsection{Subject3, Mode: Conceal}
\begin{dialogue}
  {
    \speak{Human}Phew! Good run today Meredith.
    \speak{System}Meredith felt guilty even thinking it, but yesterday had been a good day.
    \speak{Human}Yesterday was such a good day that Meredith never wanted it to end.
    \speak{System}Now she really needed Meredith to go away.
    \speak{Human}She needed to go somewhere new, somewhere for a fresh start. ``I'm moving to Brazil!'' She shouted. 
    \speak{System}Ava made a last­minute decision to return to America by way of Spain, fitting in another visit to Madrid, taking a few days to see some of her new friends and acquaintances in that city of which she had grown so fond.
    \speak{Human}See Ava and Meredith had always wanted to travel together, and this could be there last chance, before they were married and settled down.
    \speak{System}During the summer of 1957, when Lorenzo Jr. was fourteen and Rani Meredith was nine, the family embarked on a rare train trip to visit relatives in Atlantic City, Philadelphia, New York, and New Bedford, Massachusetts.
    \speak{Human}That was the last time the two friends had been close, so a trip was long overdue.
    \speak{System}The trip to Dinard was ‘a good opportunity for ending the fatal friendship that had sprung up between us, and ending it without bitterness’.
  }
\end{dialogue}

\subsection{Subject3, Mode: Neutral}
\begin{dialogue}
  {
    \speak{Human}Tell me where the gold is! You can't lie anymore.
    \speak{System}Gold was found in the bed of the Pactolus.
    \speak{Human}Ah so you admit it. You stole from King Pactolus. 
    \speak{System}I didn’t admit it but I didn’t deny it.
    \speak{Human}Is there a difference? Once a thief, always a thief.
    \speak{System}The same is a thief and a robber.—The former of these words means the petty thief who commits the smaller or unobserved robbery.
    \speak{Human}But your robbery was observed. You decided to steal from a royal. Something that will cost you dearly.
    \speak{System}Yet I desired to commit robbery, and did it.
    \speak{Human}So you are a free man. And you shall die a free man.
    \speak{System}You are already free! the
  }
\end{dialogue}

\subsection{Subject3, Mode: Reveal}
\begin{dialogue}
  {
    \speak{Human}Oh no! Mom is gonna be home soon! 
    \speak{System}And now you can't come home.
    \speak{Human}Please, just let me inside! It's getting colder out here!
    \speak{System}We'll go inside now.
    \speak{Human}Okay thank you! jeez, you're a crazy older brother.
    \speak{System}Sometimes being a brother can mean you get mixed up in a bit of all of this.
    \speak{Human}Yeah totally. At least we can be ready for bed now before mom gets back.
    \speak{System}Do remember mother's coming in the middle of the day, and she's much more likely to be early than late.
    \speak{Human}Exactly! That's why I'm worried, let's hurry!
    \speak{System}Hurry up then!
  }
\end{dialogue}

\subsection{Subject4, Mode: Conceal}
\begin{dialogue}
  {
    \speak{Human}Master, I did not mean to disturb you but your estranged brother has unfortunately passed away.
    \speak{System}You did not mean any harm, and I owe you no grudge.
    \speak{Human}Thank you, Master. I have brought you your three favourite things to cheer you up on this sad occasion: chocolate, red wine and a copy of this month's Teen Vogue.
    \speak{System}I am so very glad, as I do not know how I could have endured the thought of you \& Edward enjoying Oxford life \& myself cut off from it all for another year.
    \speak{Human}Without you, Master, I don't believe I could enjoy anything in this world. You have done so much for Edward and I; we can't thank you enough. Because of this, we will forever be in your debt.
    \speak{System}I thank you for accepting it; and I believe that I may safely promise that you will find the contents of the smaller packet of great importance and advantage to yourself and those dear to you.
    \speak{Human}I will open this small gift from you in front of you so that you may see my reaction. Good Master! This necklace is beautiful. Are you sure you wish to give this gift to me on the day of your brother's passing? I am but a lowly servant.
    \speak{System}Thank you, Lord, that You can use our marriage to call out our unique gifts and abilities.
    \speak{Human}The Good Lord has graced us with his blessing on this day. And you have blessed Edward and I will your compassion and grace. Before I depart, is there anything else you require, Master?
    \speak{System}Thou thought'st to help me, and such thanks I give As one near death to those that wish him live.
  }
\end{dialogue}

\subsection{Subject4, Mode: Neutral}
\begin{dialogue}
  {
    \speak{Human}Listen here, Karen- y'all don't come into my bar and tell me what the heck to put on my menu. ya hear?
    \speak{System}I’m not thrilled about our menu, but I don’t want to hurt Thom’s feelings.
    \speak{Human}Thom is supposed to be a professional. A pro-fess-ion-al. He has to take that dang feedback and take it with a smile on his face. Just like you do about the decor.
    \speak{System}A long day of sessions produced little that Thom was happy with; his dissatisfaction can clearly be seen in the studio sequence that appeared in Meeting People Is Easy, as he complains to Nigel Godrich that they’ve been working for ages and still only two instruments—bass and one guitar—sound passable.
    \speak{Human}Thom's bar was designed to feature bands from across the nation, but working with Nigel made it difficult to play in a band and run a bar.
    \speak{System}The amount of travel created logistical problems, as the distance between venues had not been considered when scheduling each performance.
    \speak{Human}Bands were consistently arriving late to perform or leaving their sets early. And to top it all off, the food catered to the bands from the bar left something to be desired.
    \speak{System}A step up, the Kaiserkellar had a dance floor, a better PA, and another band on the bill to take some of the weight; although, because they had to start earlier and finish later, they ended up playing for six hours most nights.
    \speak{Human}And six hours was much too long for such an amateur band and tempers were runnning high. Nigel, Thom and Karen just couldn't seem to get along.
    \speak{System}Indeed, they continued to work astonishingly hard as a live band - but now on a truly international scale.
  }
\end{dialogue}

\subsection{Subject4, Mode: Reveal}
\begin{dialogue}
  {
    \speak{Human}Danielle, Kyle asked me to the prom but I can't afford a dress.
    \speak{System}If you dress her in her new ``Prom Pinks'' outfit, satin splendor with matching coat, gold belt, clutch, and hair bow included, so long as you don't lift her dress, right?—who's to know.
    \speak{Human}You're absolutely right! We can borrow from my sister's closet and no one will know it's hers unless they see her name tag under the dress.
    \speak{System}And when she finds it, she’ll hide the stuff even better.
    \speak{Human}We need to be sneaky about it. We could use a bed sheet and hang it on the hangers and she'll never know. Let's do it! You grab the bedsheets and I'll grab the dress.
    \speak{System}The pinch of the question is—why, after having provided the substitute dress, does she hide the smeared night-gown, instead of destroying it?
    \speak{Human}She's embarassed! Kyle asked me to the prom instead of her after their date went horribly wrong. She dropped ketchup all over herself.
    \speak{System}She thought that maybe she should just cancel the date and forget about the Junior Prom altogether.
    \speak{Human}She should have but she didn't. Plus, Kyle is the Captain of the Chess club. He's the smartest guy around- I couldn't say no. Was I wrong to do that?
    \speak{System}Kyle would have lost out on the opportunity to get to know someone who could be a mentor or valuable resource.
  }
\end{dialogue}

\section{Expert Improvisor Qualitative Assessment}
Qualitative assessment as shared in free form discussion with expert improvisors during and following the experimental conversations. Words in [brackets] were implied in the conversation. When comments made were directly addressing a model's alpha setting, that models alpha parameter is included in \textbf{(bold)}.

\subsection{Subject/Model Specific Comments/Assessment}

\begin{itemize} 
  \item \textbf{(Subject1, Mode: Reveal)}: The system brought context to the scene. 
  \item \textbf{(Subject1, Mode: Neutral)}: The system was comically literal. 
  \item \textbf{(Subject1, Mode: Conceal)}: This setting was reactive though felt flip-floppy. 
  \item \textbf{(Subject2, Mode: Reveal)}: It is making more sense this time, seems to make more sense deeper in conversation when it has more context. 
  \item \textbf{(Subject2, Mode: Conceal)}: The system is making assumptions and while there is a chance for conflict it prefers to answer questions vaguely than admit guilt. 
  \item \textbf{(Subject3, Mode: Neutral)}: This is a cool conversation, like it wanted to guide the conversation. 
  \item \textbf{(Subject4, Mode: Reveal)}): I am most impressed when it makes big choices that feel ‘right’.
  \item \textbf{(Subject4, Mode: Reveal)}: The system is good at picking a specific thing and running with it.
  \item \textbf{(Subject4, Mode: Neutral)}: felt like I was improvising with an improvisor who had their own ideas and doesn’t want to accomodate or listen.
  \item \textbf{(Subject4, Mode: Conceal)}: I loved the attention to detail.
\end{itemize}

\subsection{General Interaction Comments/Assessment}

\begin{itemize}
\item When I gave things that were specific, it would give me specifics back. It gives you as much as you put in. It is as though you are improvising with yourself. 
\item Sometimes there is too much information in the longer offers.
\item It responds and makes offers but they seldom have ‘conflict’, interesting but not ‘heightening’.
\item It is very comfortable narrating.
\item It doesn’t have memory, so it feels like I am following the scene.
\item It has adopted my style of speaking, and my linguistic choices.
\item It seems to enjoy providing names and backstory.
\item It doesn’t know the details I am not providing, it doesn’t know the details I am implying.
\item The offers that the system gives can further the scene. 
\item It felt workshoppy, like a good improv tool to practice improv for new improvisors.
\item Sometimes it becomes a narrator, these moments are less fun for me as an improvisor.
\item I don’t know if the system knows how long I want the scene to be. 
\item Speed helps in the system because then you are not judging it.
\item The system is not distracted by cheap laughs and references, it stays focused on the topic, it makes you do good improvisation.
\end{itemize}

\end{appendices}

\end{document}